# Studies on DC transport and terahertz conductivity of granular molybdenum thin films for microwave radiation detector applications

Shilpam Sharma[1*], Ashish Khandelwal[1], Edward Prabu Amaladass[2,3], S. Abhirami[2,3], SK Ramjan[1,5], J. Jayabalan[4,5], Awadhesh Mani[2,3] and M. K. Chattopadhyay[1,5]

[1]*Free Electron Laser & Utilization Section, Raja Ramanna Centre for Advanced Technology, Indore 452013, India.*
[2]*Condensed Matter Physics Division, Materials Science Group, Indira Gandhi Centre for Atomic Research, Kalpakkam 603102, India.*
[3]*Homi Bhabha National Institute, IGCAR, Kalpakkam 603102, India*
[4]*Nano Science Laboratory, Materials Science Section, Raja Ramanna Centre for Advanced Technology, Indore 452013, India.*
[5]*Homi Bhabha National Institute, Training School Complex, Anushaktinagar, Mumbai 400094, India.*

\* shilpam@rrcat.gov.in

Abstract

The morphological, transport, and terahertz optical properties of DC magnetron sputtered granular molybdenum thin-films with nano-grains embedded in an amorphous molybdenum/molybdenum oxide matrix have been studied in their normal and superconducting states. The superconducting transition temperatures of these films are much higher than that of bulk molybdenum. The optical properties of these thin-films have been studied using terahertz time-domain spectroscopy. Their properties have been compared with the existing materials used for the development of radiation detectors. The films' resistivity lies in >100 µΩ-cm range, ideal for making highly sensitive radiation detectors. The Hall measurements indicate holes as the dominant carriers with very small mean free path and mobility. In the normal state, the films are disordered bad metals. However, they have large superfluid density and stiffness in their superconducting state. The properties of the films in the normal and superconducting states are promising for their use in cryogenic radiation detectors for microwave, terahertz, and far IR frequency ranges.

# 1. Introduction

Studies on the granular and homogeneously disordered superconductors have been an active area of research [1-3]. These materials exhibit exotic phenomena like the occurrence of disorder driven superconductor to insulator transition (SIT) [1,4,5] and the existence of pseudo-gap in the otherwise *s*-wave superconductors [6-8]. Recently, with the emerging possibilities of their use as the active layer of radiation detectors for microwave, terahertz, and far-infrared wavelengths, the field has also gained significant popularity in the domain of applied research [9-12].

Near the mobility edge, the granular and homogeneously disordered superconducting films undergo a continuous quantum phase transition from a coherent many-body ground state of delocalized Cooper pairs to localized incoherent quasi-particle states [13]. Strongly disordered quasi-two-dimensional superconducting films may even feature a superconducting gap in the insulating state [8]. The SIT in such films may occur with or without going through an intermediate quantum-corrected metallic ground state of incoherent but delocalized quasi-particles [1,14-16]. Strong disorder in the superconductors can suitably modify their electronic, optical, and thermodynamic properties, thereby making them apt for superconducting micro-resonator based detectors and superconducting nanowire single-photon detectors (SNSPD) [11]. These detectors find their utility in quantum computing applications. By controlling the disorder, the superconducting transition temperature ($T_C$) can be suitably tuned to detect different radiation frequencies across the spectrum. High normal state resistivity, either due to low density or localization of carriers, is a hallmark of the disordered state of otherwise good metal. The high normal state resistivity of the active detector layers is desirable as it helps in efficient photon absorption and large kinetic inductance [9].

Presently, the development of sensitive microwave kinetic inductance detectors (MKIDs) and SNSPDs is mostly based on the superconducting nitrides of Nb, Ti, or Mo [9,17]. These compounds are usually deposited as thin-films using reactive physical vapor deposition (PVD) techniques. The reactive deposition offers excellent control over the properties of the superconducting compounds but the process itself suffers from target poisoning and hysteresis in the deposition. Thus, to ease the detector fabrication, elemental superconductors' use as the active detector layers is more convenient. Except for the use of granular aluminum films [10], there is hardly any report on the utilization of disordered films of elemental superconductors for such applications. Molybdenum (Mo) is one element that finds its extensive use in thin-film based particle detection at cryogenic temperatures [18]. Its stability at high process temperatures, ability to form low resistivity Ohmic contacts, and insolubility with metals like Cu and Au, make Mo an indispensable material for the microelectronic device industry. In addition to being device fabrication-process friendly, its normal state resistivity and the $T_C$ tunability from hundreds of mK to about 8 K by controlling the disorder [19] make it an excellent choice for developing other types of detectors too.

Mo is a refractory metal, and thus its thin-films can be deposited using PVD techniques like electron-beam evaporation [20], ion beam, and magnetron sputtering [21,22]. Among the different PVD techniques, magnetron sputtering is a relatively simpler and efficient technique where a broad set of process parameters can be tuned to deposit films with desired properties. The process can easily be scaled to deposit large surface area films and coatings for industrial applications. We have previously reported the effect of argon pressure and negative substrate bias on the morphological and electrical properties of DC sputtered Mo films [14,23]. The deposition rate and argon pressure during the deposition can reproducibly tune the grain sizes and disorder in the films [14]. Unlike many other systems, the disorder driven SIT in granular Mo films occurs via a bad metallic state [14]. The optical response of disorderd Mo films at microwave and THz frequencies needs to be characterized to ascertain their usefulness for microwave resonator applications.

Here we report on the DC transport and terahertz (THz) optical conductivity of granular Mo thin-films with nano-granular morphology and high DC resistivity. Based on our previous studies on the Mo films[14,23], the deposition parameters have been selected to deposit the thin film samples having residual resistivity ratios (RRR = $R_{295K}/R_{1.1Tc}$) ~1. Such samples have been previously shown to remain in a state with high resistivity but still far from the localized insulating state [14]. The present samples have been characterized for phase purity, grain morphology, DC electrical transport, and their optical response in the THz frequency range. To the best of our knowledge, the THz optical properties of the disordered Mo films in the normal and superconducting states are, hitherto, not documented. The systematic characterization of the morphology, DC transport, and THz optical properties is essential to determine the usefulness of these films for the MKID and SNSPD applications. The estimation of the normal state charge carrier density and mobility, sheet resistance, superconducting gap, superfluid density, and penetration depth is of utmost importance as the sensitivity of MKIDs depends upon these properties of the films.

## 2. Experimental details

### 2.1. Thin-film preparation, characterization, and electrical transport measurements

Granular films of molybdenum (Mo-A and Mo-B) were deposited on $SiO_2$ (300 nm thick, amorphous) coated Si (100) substrates (0.5 mm thick) by DC magnetron sputtering using high purity (99.95 %) Mo target (Testbourne Ltd., UK) and ultra-high purity argon gas (99.9995 %). The sputter deposition system is equipped with three 1″ magnetron guns. The magnetron cathodes are arranged in planetary configuration, and the target to substrate distance can be varied from 50- 110 mm. Before their loading in the deposition system, the 10 mm × 10 mm substrates were cleaned ultrasonically in boiling acetone followed by rinsing in de-ionized water, rinsing in ethyl alcohol, and were finally blow-dried. The deposition setup is equipped with a load lock system for substrate manipulation inside and out of the deposition chamber maintained at ultra-high vacuum (better than $1\times10^{-7}$ mbar). The thin film samples were deposited under the background argon pressure of 3.2 µbar (Mo-A), and 3.4 µbar (Mo-B) with the substrate kept at 75 mm from the target. The residual impurity gases (primarily $O_2$ and water vapor) and the deposition rate have an immense bearing on the films' morphology and electrical properties [14,23]. The deposition system can be pumped to an ultra-high vacuum. However, to increase the background impurities of $O_2/H_2O$, we have pumped the system to only a high vacuum ($10^{-6}$ mbar) range. The impurity impingement rate in a chamber pumped to $10^{-6}$ mbar contributes roughly one impurity monolayer in less than a minute. This large impurity flux and low power sputtering were used to control the disorder and synthesize samples with large normal state resistivity and RRR ~1. These properties are manifestations of the samples' physical form, viz., nano-grains embedded in the amorphous Mo/MoOx matrix. The UHV system was allowed to be pumped only up to $2\times10^{-6}$ and $1.7\times10^{-6}$ mbar base pressures for Mo-A and Mo-B films, respectively. To achieve deposition rates comparable to the impurity impingement rate, the sputtering was performed with a small constant current of 40 mA and voltage varying from 314 to 325 V, thereby delivering approximately 13 W power onto the target. Prior to the deposition of the films, the Mo sputtering target was pre-cleaned by sputtering for 10 minutes at 13 W. This low deposition rate rendered the films granular with nano-grains embedded in the amorphous Mo/MoO$_X$ matrix.

The phase purity of the two films has been confirmed using grazing incidence X-ray diffraction (GIXRD) measurements using an Equinox 2000 x-ray diffractometer (Inel, France) employing Cu-K$\alpha$ radiation. The GIXRD data has been collected using a curved position sensitive detector with the angle of incidence kept fixed at 2º. The film's thickness has been measured using X-ray reflectivity measurements (XRR) performed using a Bruker, GmbH make D8 diffractometer. To ascertain the films' surface morphology and grain sizes, the atomic force microscopy (AFM) measurements were performed in a Bruker MultiMode 8-HR AFM, GmbH system. The films' electrical properties as a function of temperature varying from 1.7 to 300 K have been measured in a magneto-optical cryostat (Cryofree Spectromag CFSM7T-1.5) from Oxford Instruments, UK. The DC electrical resistivity and sheet resistance have been measured in the linear four-probe geometry. The electrical contacts on the samples were made using 25 μm thick copper wire attached to the samples using high conductivity silver paint. The Hall effect measurements have been performed at different temperatures using the AC transport (ACT) option of a 9 T physical properties measurement system (PPMS, Quantum Design, USA).

## 2.2. Terahertz time-domain spectroscopy

Temperature-dependent time-domain spectroscopy in 0.2 to 1.1 THz frequency range was performed in a custom-built THz time-domain spectrometer (Teravil Ltd./ Ekspla uab, Lithuania) in the standard transmission geometry. The spectrometer is built around the Cryofree Spectromag CFSM7T-1.5 cryostat mentioned above. The windows on the outer vacuum jacket and the variable temperature insert (VTI) of the cryostat are made of Z-cut quartz. The output of a mode-locked diode-pumped solid-state laser (1030 nm, 96 fs pulse duration, and 76 MHz repetition rate) is split into excitation and detection pulses. The excitation pulses are focused on the gap between the electrodes of a Bi-doped GaAs based photoconductive antenna. The emitted THz pulses are coupled out of the antenna using an integrated collimating lens (Si). With the help of two off-axis parabolic mirrors, the THz pulses from the photoconductive antenna are focused at the center of the sample space inside the cryostat. The time-dependent electric field of the transmitted THz pulses is detected using a similar GaAs photoconductive antenna-based detector. The detector is irradiated by the detection laser pulse, which is delayed by tens of picoseconds with respect to the excitation pulses. The cryostat is fitted with a motorized sample insert that brings the thin film sample and the reference substrate (amorphous $SiO_2$ coated Si substrate) to the focus of the THz beam of 4 mm diameter one after the other, at each temperature. The electric field amplitude of the direct beam is also recorded at each temperature. The multiple reflections arising from the substrate and the cryostat windows have been removed by appropriately setting the initial delay of the measurement.

The measured THz transient electric field is Fourier transformed using an FFT routine to yield the spectral information (amplitude and phase) out of the signal transmitted through the sample, reference substrate, and the direct beam. To delineate the film's optical properties from that of the substrate, the sample's experimental complex transmission function is calculated as the ratio of transmissions through the sample to the reference substrate. Simultaneously, the complex transmission function for the reference substrate is calculated as the ratio of transmissions through the reference to the direct beam. The theoretical complex transfer functions for both the sample and the reference are computed using the Fresnel equations [24]. To compute the complex optical constants, the error between the experimentally calculated and theoretical transfer functions is minimized at each frequency using the Nelder-Mead simplex algorithm [25] implemented in Python script [26]. The refractive index ($n$) and extinction coefficient ($k$) of the reference have been estimated at each frequency. They are fed as parameters for the optimization of the optical constants of thin films.

## 3. Results and discussion

### 3.1. Crystal structure and surface morphology

The deposition chamber's base pressure and argon's pressure during deposition have been selected so that the deposited films have nano-crystalline grains in a disordered (amorphous Mo and MoOx) matrix. The growth of films having RRR ~1 and high sheet resistance have been aimed for. GIXRD measurements have been performed to ascertain the phase purity of the Mo thin films. The plots of normalized intensity as a function of diffraction angle ($2\theta$) are shown in the Figs. 1(a) and (b) for the samples Mo-A (deposited at $2\times10^{-6}$ mbar base pressure and 3.2 µbar argon pressure) and Mo-B (deposited at $1.7\times10^{-6}$ mbar base pressure and 3.4 µbar argon pressure) respectively. All the reflections observed in the GIXRD patterns have been indexed to the body-centered cubic (*bcc*) Mo *d*-spacing [27]. A small peak from the underlying Si substrate is also observed in the plots. The X-ray data have been refined using the PowderCell for Windows software [28], and the fitted profile is shown superimposed on the raw data in Figs. 1(a) and (b). The value of the lattice parameter '*a*' for the samples Mo-A and Mo-B are 3.1469 and 3.1465 Å, respectively. These lattice parameter values are slightly smaller than the reported value of 3.1472 Å [27], which is probably due to the compressive stress in the nano-meter sized grains in these films.

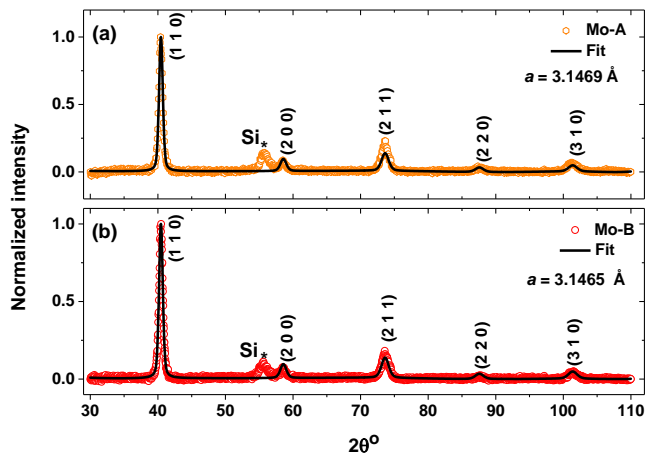

**Figure 1(a):** GIXRD pattern of molybdenum thin film deposited at 3.2 µbar pressure showing all the reflections due to the *bcc* Mo *d*-spacing. A small peak near 55° due to the reflections from the Si (100) substrate has been marked with '*.' The fit to the data (black line) is also shown superimposed on the experimental data (red open circles). **(b):** GIXRD pattern for the Mo film deposited at 3.4 µbar pressure also shows similar peak positions and lattice constant value.

The precise measurement of the film thickness is vital for the accurate estimation of their optical properties. The films' thickness and roughness have been estimated by fitting the XRR pattern using the Parratt32 routine [29]. The XRR pattern and the fitting for the Mo-A and Mo-B films are presented in Figs 2(a) and (b), respectively. From these measurements, the Mo-A film was found to be around 22.4 nm thick with 0.59 nm surface roughness, while the Mo-B film was around 24.3 nm thick with 0.78 nm surface roughness.

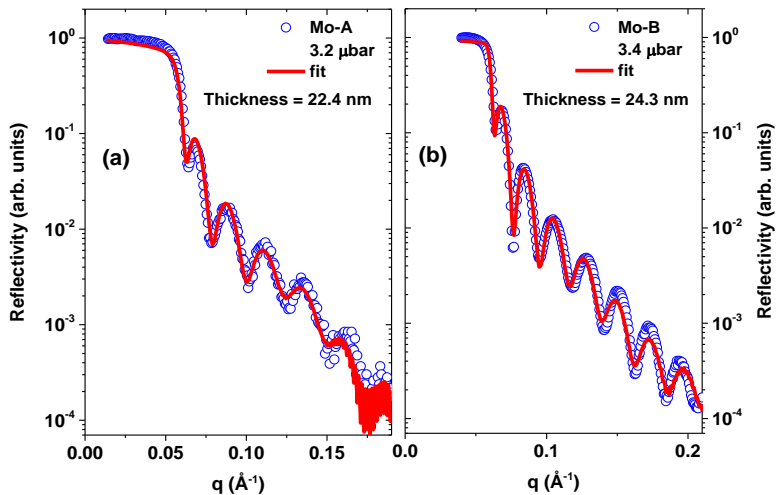

**Figure 2(a):** X-ray reflectivity (XRR) as a function of the scattering vector for the Mo-A thin film sample. The film's thickness and roughness have been estimated by fitting the data to a 22.4 nm thick film model. (**b**): XRR data along with the fit for the Mo-B film. The film thickness has been estimated to be 24.3 nm.

The topography of both films has been studied using AFM. The two- and three-dimensional micrographs are shown in Figs. 3(a)- (b) for Mo-A and in Figs. 3(d)- (e) for Mo-B. The surface morphology of the films gets affected by the residual impurities in the deposition chamber. It can be seen from the micrographs that the thin film Mo-A deposited with the chamber pumped to higher base pressure ($2\times10^{-6}$ mbar) has less number of grains as compared to sample Mo-B deposited with a somewhat better base vacuum ($1.7\times10^{-6}$ mbar). It can be observed from the three dimensional (3D) micrographs that the coalescence of the nano-crystallites with obscure boundaries in Mo-A has given rise to the elongated grain morphologies. On the other hand, the sample Mo-B has roughly spherical grains with sharp boundaries and homogeneous distribution. Similar surface morphologies have been reported previously for the sputter-deposited Mo films [23,30-32].

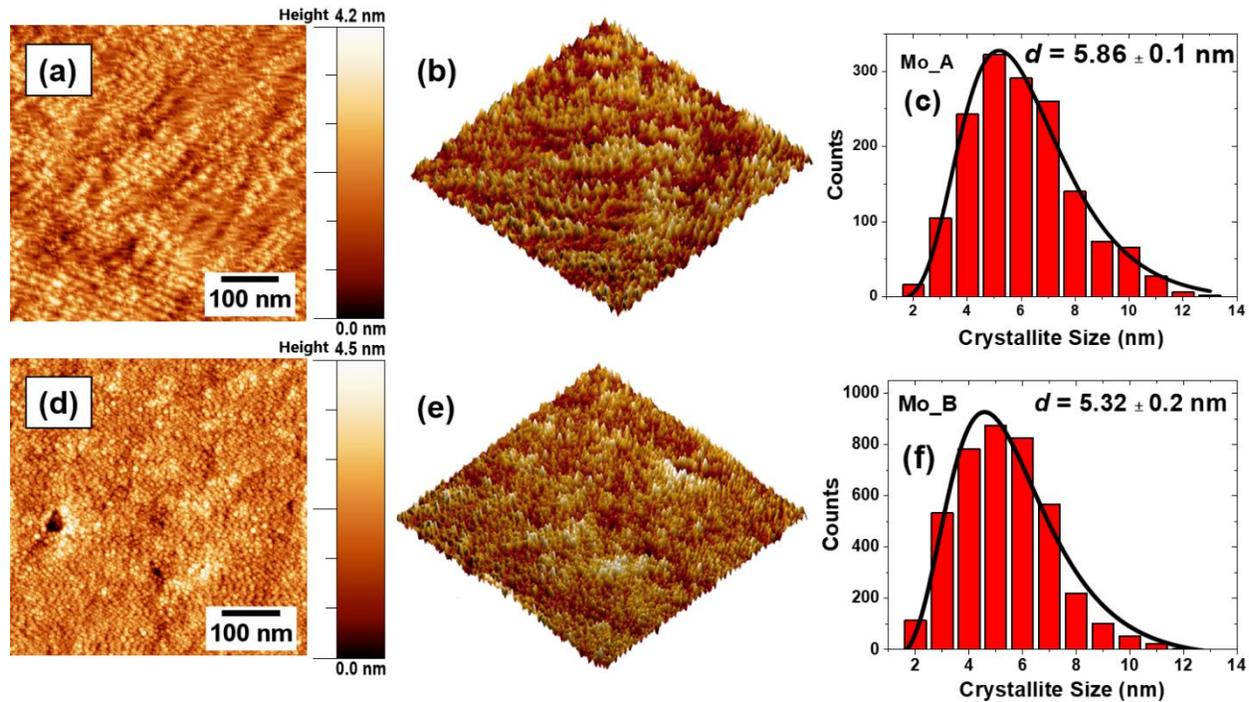

**Figure 3 (a):** Two and **(b):** three dimensional AFM micrograph of the Mo-A thin film showing nano-crystallites with obscure boundaries. The crystallites are arranged in elongated geometries. The RMS surface roughness of the film is estimated to be 0.55 nm. (**d**) and (**e**): Two- and three-dimensional AFM micrograph of the Mo-B thin film sample indicating the growth of nano-grains with sharp boundaries and homogeneous distribution. The color bars represent the variation in the height of the grains. The film's RMS surface roughness is 0.59 nm **(c)** and **(f):** Size distribution of the crystallites along with the log-normal fit for Mo-A and Mo-B thin films, respectively. The average crystallite size of the two samples does not vary much, but the Mo-A sample's grain density is much lower than that in the Mo-B sample.

The effect of impurities on Mo thin films' morphology and physical properties has already been reported [33,34]. For metals like Mo with atoms having low surface mobility, background impurities like $O_2$ and water vapor further hinder the grain growth during film deposition. For this reason, the sample deposited under better vacuum conditions (Mo-B) has a higher number of grains with sharp grain boundaries. Though for both the samples, the gaseous background impurities might have been equally hindering the growth of large-sized grains, the slightly higher $O_2$ partial pressure during the deposition of Mo-A films has led to fewer grains in comparison with the amorphous oxide matrix. The root mean square (RMS) surface roughness of the films has been estimated from the AFM micrographs using the WSXM software [35]. The films' RMS surface roughness has been estimated to be 0.55 nm for Mo-A and 0.59 nm for Mo-B. The RMS surface roughness value is close to the roughness estimated from the XRR measurements presented above. The Mo films with similar small RMS surface roughness have been reported earlier as well [23,36]. Mo films' low surface roughness is desirable for optical measurements as it reduces the diffused scattering from the surface asperities, thereby allowing better THz transmission. From the device fabrication point of view, the low RMS surface roughness of the active layer facilitates the formation of low resistance Ohmic contacts with the electrode materials required for the electrical biasing of the device. This also reduces the electrical noise and heat generation in the detectors operating usually at the sub-Kelvin temperatures.

The average lateral grain sizes have been estimated from AFM micrographs using the NT-MDT SPM image processing software and ImageJ software [37]. The size distributions of approximately 1500 and 4000 grains from the whole micrographs of Mo-A and Mo-B samples respectively have been fitted to the log-normal distribution. The average lateral grain sizes have been estimated to be $5.8 \pm 0.1$ nm and $5.3 \pm 0.2$ nm for the Mo-A and Mo-B film respectively. The grain size distribution of these films and the respective log-normal fits have been presented in Figs. 3(c) and (f), respectively. It can be noticed from the size distribution plot that the average grain size is roughly the same for both the thin films. Though the grain sizes do not vary much, the number of grains is appreciably lower in Mo-A than in the Mo-B film. This reduced number of grains in the Mo-A sample can be attributed to the enhanced growth of amorphous Mo or molybdenum oxide, mainly due to the higher partial pressure of the deposition system's background impurities. The grain density has a strong bearing on the transport and optical properties of the thin film samples. The electrons get knocked at the grain boundaries, or the amorphous inter-grain regions hinder their motion by reducing the hopping probability, thereby increasing the resistivity[38].

## 3.2. Electrical transport properties

The temperature dependence of the Mo films' electrical resistivity in the temperature ranges 2- 6 K and 6-300 K are shown in Figs. 4(a) and (b), respectively. It can be seen from Fig. 4(a) that both the samples show a superconducting transition to a zero resistivity state at a temperature much higher than the superconducting transition temperature ($T_C$) of the bulk Mo (~0.92 K).

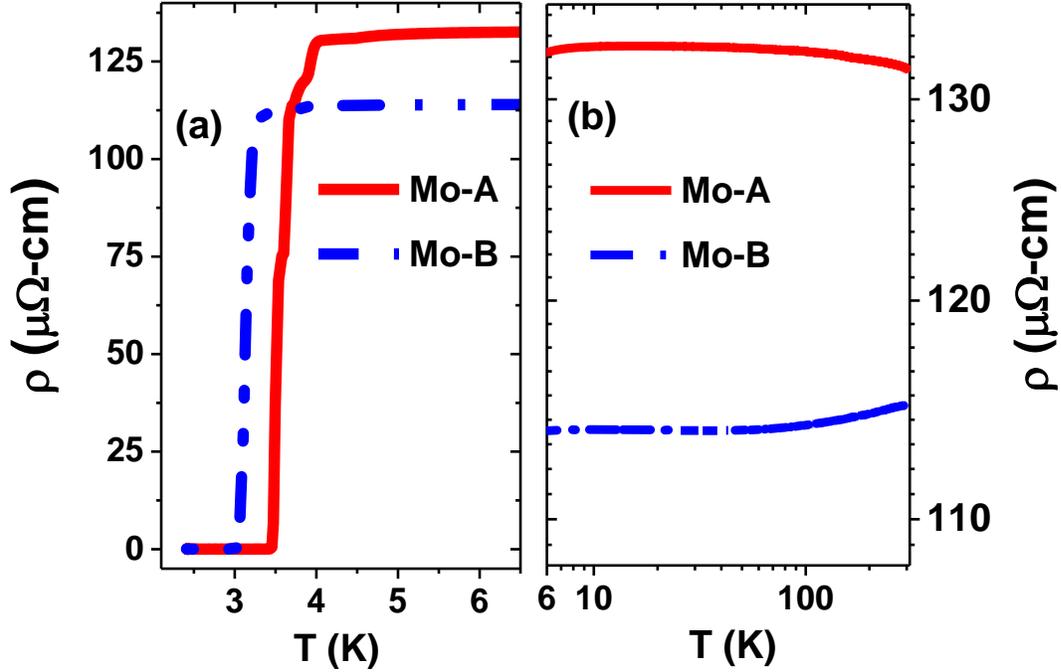

**Figure 4(a)**: Resistivity as a function of temperature for the Mo-A and Mo-B samples. The superconducting transition occurs at a temperature (4.17 K for Mo-A and 3.23 K for Mo-B) much higher than the bulk Mo $T_C$. **(b):** log-log plots of the normal state resistivity as a function of temperature for both the thin films. At high temperatures, the Mo-A film shows a negative temperature coefficient of resistivity (TCR), and the Mo-B film shows positive TCR. Below 100 K, the resistivity of both films becomes temperature independent. The residual resistivity ratio (RRR) of both the samples remains close to 1.

The Mo-A film becomes superconducting at 4.17 K, and the global $T_C$ for the Mo-B film is at 3.23 K. The increase in the $T_C$ of the films may be ascribed to the surface phonon softening due to increased surface to volume ratio, thereby increasing the electron-phonon coupling strength [39]. It can also be caused by the increased spectral density around the Fermi surface due to fluctuations in the discrete energy levels of the nano-crystalline grains [40,41]. Similar enhancements in the $T_C$ have already been reported earlier [14,42]. The log-log plot in Fig. 4(b) shows the normal state resistivity variation as a function of temperature for the Mo-A and Mo-B samples. At high temperatures, the Mo-A thin film shows a negative temperature coefficient of resistance (TCR) while the Mo-B film has positive TCR. It can be observed in Fig. 4(b) that below 100 K and down to superconducting transition, the resistivity of both the samples remains nearly independent of temperature. The residual resistivity ratio (RRR = $R_{295K}/R_{1.1Tc}$) has been estimated to be 0.99 and 1.01 for Mo-A and Mo-B, respectively. The room temperature resistivity ($\rho_{300K}$) of the Mo-A film is

131.6 µΩ-cm, and its sheet resistance ($R_S$) is 58.7 Ω/□. For the Mo-B film, the $\rho_{300K}$ is 115.1 µΩ-cm, and the $R_S$ is 47.5 Ω/□. These resistivity values are two orders higher than the bulk Mo resistivity, which is ~5.4 µΩ-cm. Hence, both the samples have high room temperature resistivity, high sheet resistance, and RRR ~1. The sample deposited under better base vacuum (Mo-B) shows better grain connectivity and positive TCR while it is negative for the slightly more disordered Mo-A sample. For the films with thickness less than the penetration depth, the kinetic inductance $L_K$ is proportional to the $R_S$, following the relation $L_K = \frac{\hbar R_S}{\pi \Delta_0}$, where $\Delta_0$ is the superconducting gap at zero temperature [9]. Thus, the high values of resistivity and $R_S$ help achieve large kinetic inductance in highly compact thin-film devices. Highly sensitive MKIDs fabricated using TiN thin films with similar $\rho_{300K}$, $R_s$, and RRR have already been reported [9].

The Hall effect measurements have been performed on the samples at different temperatures from 50- 300 K. Mo is a compensated metal with the mobility of holes higher than that of the electrons ($\mu_h$ ~ $2\mu_e$ at 300 K) [43] giving rise to a positive Hall coefficient. Figure 5 presents the results of Hall measurements. The transverse resistivity as a function of the magnetic field for both the samples at 300 K is shown in the inset of Fig. 5(a). The slope of this curve is the Hall coefficient ($R_H$), which is positive for both the samples. Hole-type carriers dominate the transport in the samples. The variation of charge carrier density ($n$) as a function of temperature has been presented in the main panel of Fig. 5(a). The $n$ of Mo-A film is slightly higher than that of Mo-B, and the carrier density largely remains temperature independent below 200 K for both the films. At 300 K, the $R_H$ ~ $1.3 \times 10^{-11}$ m³C⁻¹ and $n$ ~ $4.6 \times 10^{29}$ m⁻³ for the sample Mo-A, while $R_H$ ~ $1.5 \times 10^{-11}$ m³C⁻¹ and $n$ ~ $4.0 \times 10^{29}$ m⁻³ for Mo-B. Similar values of $R_H$ and $n$ have already been reported for disordered Mo films [14,44]. The mobility of the samples has been calculated using the DC electrical resistivity ($\rho_{300K}$) and $R_H$ as $\mu = R_H/\rho$. The mobility values along with the Fermi wave vector $k_F = (3\pi^2 n)^{\frac{1}{3}}$ have been utilized to estimate the mean free paths using the relation $l_e = (\hbar\mu/e)k_F$, where $e$ is the electronic charge. It can be observed in Fig. 5(b) that over the whole temperature range of measurement, the mean free path of charge carriers in the Mo-A sample is lower than that of Mo-B.

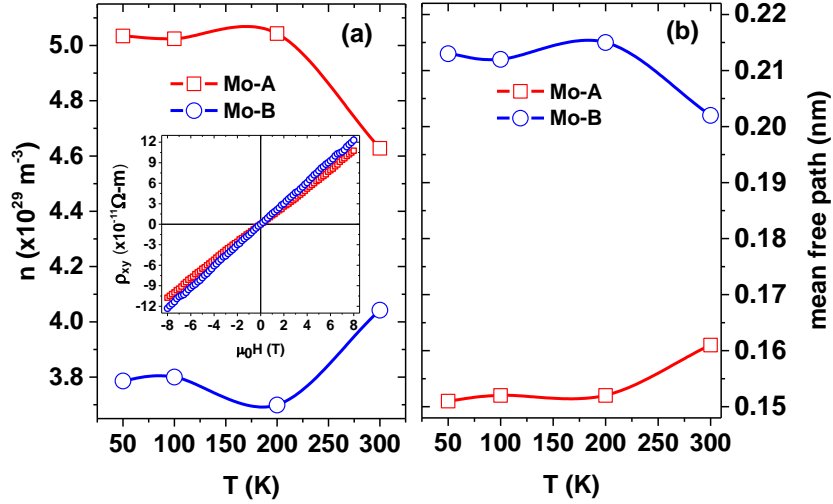

**Figure 5(a)**: The temperature dependence of charge carrier density for the Mo-A and Mo-B thin films. The carrier density of the Mo-A film is marginally higher, though the order of carrier density is same for both the films, at all the measured temperatures. **Inset:** Hall resistivity as a function of the magnetic field indicating a dominant role of the hole-type charge carriers. **(b):** Mean free path of both the samples as a function of temperature. The mean free path of the Mo-A sample is lower than that of Mo-B.

At room temperature, the carrier mobility and mean free path for the Mo-A film have been estimated to be $1.03\times10^{-1}$ cm$^2$/V-s and $1.6\times10^{-10}$ m. Simultaneously, for Mo-B the values are $1.34\times10^{-1}$ cm$^2$/V-s and $2.0\times10^{-10}$ m. Despite the high carrier density, the carriers are tending to localize due to very small values of mobility and mean free paths as compared to those of bulk Mo [43,45]. This reduction of $l_e$ in the granular Mo films has been reported earlier as well [14] and is the reason behind the poor metallic character of the films with two orders higher $\rho_{300K}$ values as compared to the bulk Mo.

In the non-interacting systems with free electrons, the Ioffe-Regel parameter $k_Fl_e$ is a unique measure of the electronic disorder. In contrast with the dielectric state, the effect of electron-electron interactions in the metallic state can be minimal at room temperature. It is for this reason that the $k_Fl_e$ for both the samples has been estimated from the $\rho_{300K}$ and the room temperature $R_H$ using the free-electron formula $k_Fl_e = \{(3\pi^2)^{\frac{2}{3}}\hbar R_H^{\frac{1}{3}}\}/[\rho e^{5/3}]$, where, $\hbar$ is Planck's constant [46]. The values of $k_Fl_e$ for the Mo-A and Mo-B films are found to be 3.8 and 4.6, respectively. These values are small, but still much higher than 1 at which the system crosses over into a weak insulating phase. Thus, the present samples are disordered metals with short mean free paths arising due to increased scattering from the amorphous inter-grain boundaries. Thin films with large $\rho_{300K}$ but well away from the SIT helps in enhancing the microwave response while still keeping the effect of quantum fluctuations and film inhomogeneity on the detector response to a minimum [10].

**3.3 Terahertz optical properties**

The three dimensional (3D) plots of complex optical conductivity ( $\hat{\sigma} = \sigma' - i\sigma''$ ) as functions of frequency and temperature for the Mo-A film, obtained from our THz time-domain spectroscopy measurements, are shown in Figs. 6(a)- (c). Figures 6(a) and (b) show the real ($\sigma'$) and imaginary ($\sigma''$) parts of the conductivity measured at temperatures from just above the $T_C$ (~1.1 $T_C$) down to 2 K. At the lowest temperature, before vanishing, $\sigma'$ shows a large dip in the low-frequency range. This is a signature of a gap opening in the quasi-particle states [47]. The 3D plots of $\sigma''$ in Fig. 6(b) are shown in the reverse temperature and frequency order for better visualization. The $\sigma''$ at $T < 1.1T_C$ shows $1/\omega$ dependence, which is typical of the superconducting state [48]. In the high-frequency range (beyond 0.6 THz) and temperatures above $T_C$, the imaginary component ($\sigma''$) becomes negative. Similar negative $\sigma''$ has already been reported for highly disordered NbN films [48]. The disorder driven localization of the normal electrons contributes to driving the normal state away from the Drude form and gives rise to this negative $\sigma''$ [48].

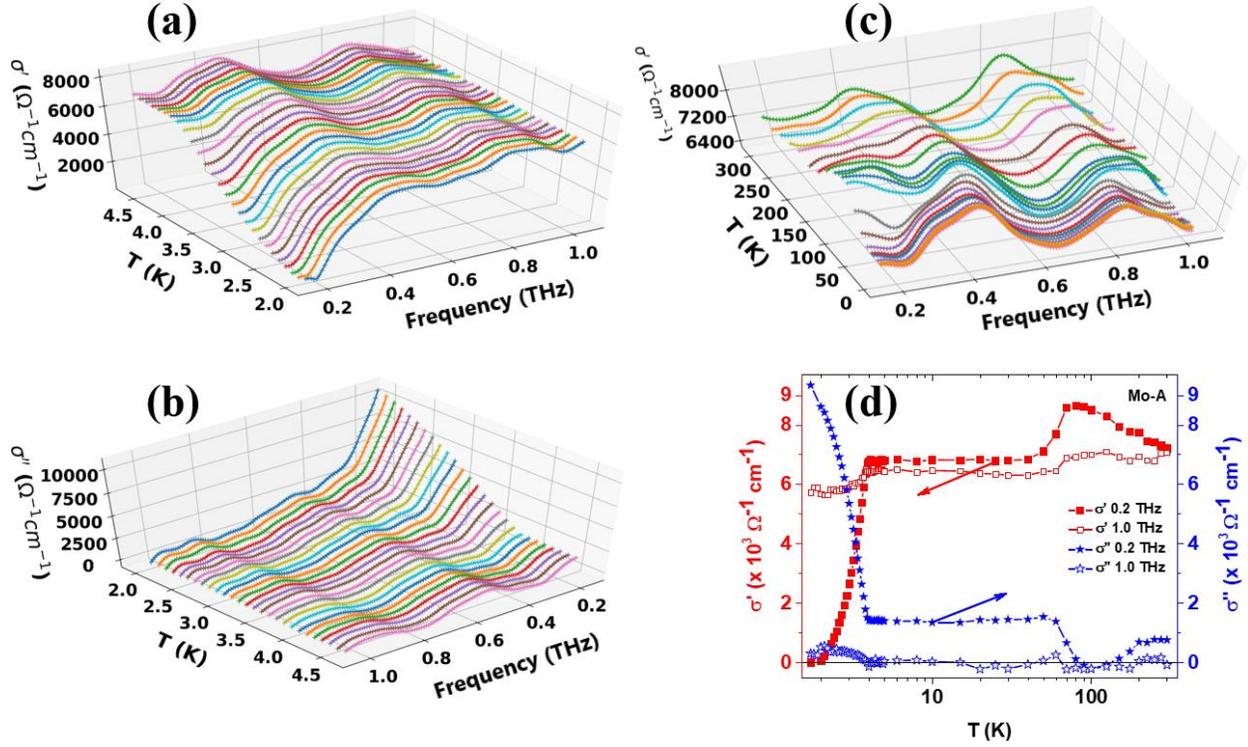

**Figure 6(a)**: 3D plots of the real part of THz conductivity of the Mo-A film. Below $T_C$ shows a dip in the low-frequency range, thereby indicating the superconducting gap in the quasi-particle states. (**b**): Frequency dependence of the imaginary part of conductivity at $T < 1.1 T_C$. The plots of the imaginary part of THz conductivity are shown in reverse temperature order for better visualization. The imaginary part of THz conductivity has a $1/\omega$ frequency dependence and is negative at higher frequencies. (**c**): The 3D plot of THz conductivity measured in the normal state of the film showing temperature and frequency independent behavior in the entire measured spectral range. (**d**): The temperature dependence of the real (red square) and imaginary (blue stars) parts of THz conductivity from 0.2 to 1 THz. The temperature dependence is consistent with the DC transport measurements presented above.

The frequency dependence of $\sigma'$ of the Mo-A film at $T > 1.1\ T_C$ is presented in Fig. 6(c). It can be observed that the normal state $\sigma'$ is mostly independent of temperature at all the measured frequencies. A broad dip in the absorption can be observed in the frequency range 0.6– 0.8 THz at all temperatures. The second dip at ~0.3 THz can be seen in the $\sigma'$ up to a temperature of 100 K, beyond which it disappears. The temperature dependence of $\sigma'$ and $\sigma''$ at the two extreme frequencies (0.2 THz and 1.0 THz) is shown in Fig. 6(d). The film shows the signatures of the superconducting transition, even at 1.0 THz. The temperature-independent behavior and the low frequency (0.2 THz) value of σ' at room temperature (7307.2 $\Omega^{-1}\text{cm}^{-1}$) is similar to our measured DC conductivity ($1/\rho_{300K} = 7598.8\ \Omega^{-1}\text{cm}^{-1}$) of the Mo-A thin film [c.f.

Fig. 4(b)]. At low frequencies and just below the $T_C$, the $\sigma'(T)$ usually shows a broad coherence peak [47,49]. No such coherence peak has been observed in the $\sigma'(T)$ curve of the Mo-A film shown in Fig. 6(d). However, a broad peak in the $\sigma'(T)$ can be observed near 100 K in both 0.2 and 1.0 THz plots. The $\sigma''(T)$ also shows a broad minimum in this temperature regime for both these frequencies. The absence of any peak features in the reference substrate's complex conductivity in this temperature range (not shown here) rules out the possibility of experimental artifacts in the data. The origin of this broad peak could not be ascertained at present. However, it can be observed from the $\rho(T)$ curves in Fig. 4(b) that the otherwise increasing $\rho$ (with decreasing temperature) of the Mo-A film becomes temperature independent below 100 K and remains unaltered down to the $T_C$. In the films of some s-wave superconductors, the disorder-induced pseudo-gap has been observed up to temperatures as high as 14 times that of $T_C$ [6]. Whether the variation in the $\sigma'(T)$ and $\sigma''(T)$ near 100 K is associated with a similar pseudo-gap opening in the Mo films needs to be investigated using tunneling measurements on similar films. The spectral weight (SW) analysis has been performed using the partial sum rule, where $SW(\omega_c) = \frac{120}{\pi} \int_0^{\omega_c} \sigma'(\omega) d\omega$, with $\omega_c$ as the cut off frequency taken as 1 THz. The SW is related to the superfluid density ($n_s$) through the relation $SW = 4\pi n_s e^2/m$. At 2K, the $n_s$ for the Mo-A film is $4.55 \times 10^{29}$ m$^{-3}$. This value is of the same order as the charge carrier density of the Mo-A sample estimated using Hall measurements (c. f. Fig. 5(a)).

The frequency dependence of $\sigma'$ and $\sigma''$ for the Mo-B thin film in different temperature ranges are presented in Figs. 7(a)- (c). The behavior of $\sigma'$ and $\sigma''$ for the Mo-B film is similar to that of Mo-A film, except that the sample shows superconductivity at around 3.2 K. As can be seen from Fig. 7(c), the normal state $\sigma'$ is independent of temperature. Similar to that of the Mo-A film, the broad dip in $\sigma'$ at 0.6– 0.8 THz and a sharp dip at 0.3 THz can be seen in the Mo-B film as well. The low frequency (0.2 THz) value of $\sigma'$ at room temperature (8688.1 $\Omega^{-1}$cm$^{-1}$) is of the same order as measured in the DC transport measurements (8598.7 $\Omega^{-1}$cm$^{-1}$) (c. f. Fig. 4).

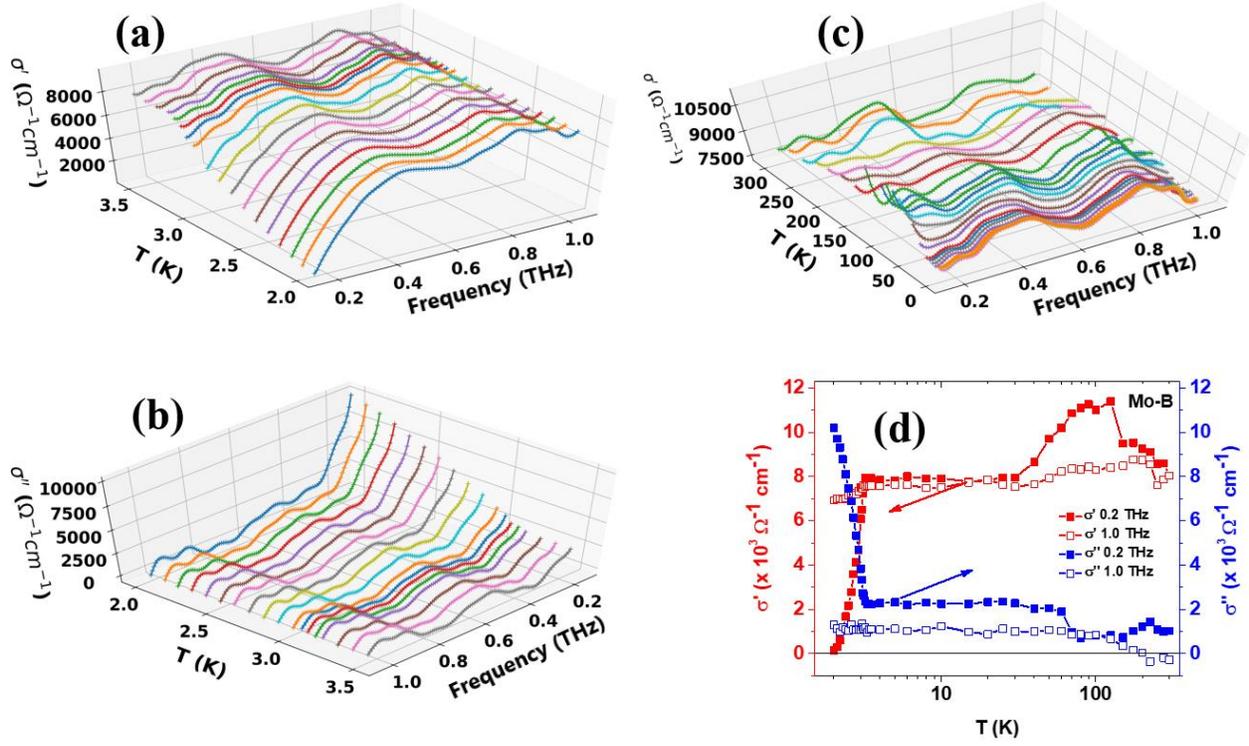

**Figure 7(a)-(b)**: Frequency dependence of real and imaginary parts of THz conductivity of the Mo-B film below the $T_C$. **(c):** Real part of conductivity as a function of frequency and temperature in the normal state of the Mo-B film. **(d):** Temperature dependence of real and imaginary parts of THz conductivity at 0.2 THz and 1 THz. The temperature dependence is consistent with the DC transport measurements.

The temperature variation of $\sigma'$ and $\sigma''$ of the Mo-B film has been plotted in Fig. 7(d) for 0.2 and 1.0 THz frequencies. Similar to the Mo-A film, the existence of the peak in $\sigma'(T)$ and an associated broad feature in $\sigma''(T)$ are observed approximately at the same temperatures around 100 K. The plot in Fig. 4(b) for the Mo-B sample also shows that $\rho(T)$ becomes nearly independent of $T$ below 100 K, while it decreases with decreasing temperature above 100 K. The features around 100 K are less pronounced in the measurements at 1.0 THz frequency.

The Mattis-Bardeen (MB) formalism describes the electrodynamics of a conventional Bardeen Cooper Schrieffer (BCS) superconducting state in the dirty limit [50]. The frequency-dependent complex optical conductivity of the present Mo films below $T_C$ is expected to follow the MB formalism. To normalize out the single-electron states' matrix elements and include the localizing effect in the spectra, the

$\sigma'$ in the superconducting state is usually normalized to the normal state conductivity just above the $T_C$ (~1.1$T_C$). The MB equations for the real and imaginary parts of optical conductivity are [50]:

$$\frac{\sigma'(\omega,T)}{\sigma_n} = \frac{2}{\hbar\omega}\int_\Delta^\infty \frac{[f(\mathcal{E})-f(\mathcal{E}+\hbar\omega)](\mathcal{E}^2+\Delta^2+\hbar\omega\mathcal{E})}{(\mathcal{E}^2-\Delta^2)^{\frac{1}{2}}[(\mathcal{E}+\hbar\omega)^2-\Delta^2]^{\frac{1}{2}}}d\mathcal{E} + \frac{1}{\hbar\omega}\int_{\Delta-\hbar\omega}^{-\Delta} \frac{[1-2f(\mathcal{E}+\hbar\omega)](\mathcal{E}^2+\Delta^2+\hbar\omega\mathcal{E})}{(\mathcal{E}^2-\Delta^2)^{\frac{1}{2}}[(\mathcal{E}+\hbar\omega)^2-\Delta^2]^{\frac{1}{2}}}d\mathcal{E} \qquad (1)$$

$$\frac{\sigma''(\omega,T)}{\sigma_n} = \frac{1}{\hbar\omega}\int_{\Delta-\hbar\omega,-\Delta}^{\Delta} \frac{[1-2f(\mathcal{E}+\hbar\omega)](\mathcal{E}^2+\Delta^2+\hbar\omega\mathcal{E})}{(\mathcal{E}^2-\Delta^2)^{\frac{1}{2}}[(\mathcal{E}+\hbar\omega)^2-\Delta^2]^{\frac{1}{2}}}d\mathcal{E} \qquad (2)$$

For $\hbar\omega > 2\Delta$, the lower limit of the integral in equation (2) becomes $-\Delta$. Here $f(\mathcal{E},T)$ is the Fermi-Dirac distribution function, $\Delta$ is the superconducting gap, $\varepsilon$ is the energy, and ω is the frequency. The plots of the normalized real part of optical conductivity at different temperatures below 1.1$T_C$ are shown in Figs. 8(a) and (d) for the Mo-A and Mo-B films, respectively. For $T$ = 2 K, the $\sigma'(\omega)$ for both the samples vanishes at the optical gap edge around 0.2 THz. The plots of the normalized $\sigma'$ and $\sigma''$ for the sample Mo-A at 2 K are shown in Figs. 8(b) and (c), respectively, along with the fitted lines obtained using the MB-formalism (eq. 1 and 2, shown as solid lines) [50]. The 2$E_g$ value for the Mo-A sample has been estimated to be 0.85 meV. The MB formalism considers a constant density of states (DOS) for normal electrons and a BCS like DOS below $T_C$. Still, in the disordered systems, the DOS can have spatial variations on the scale of the superconducting order parameter [51]. The effect of these spatial variations in the DOS normalizes out in $\sigma'$ [51], and thus the MB formula fits the data in Fig. 8(b) reasonably well. The effect of variation in the DOS and the large scattering rate is more pronounced in the case of $\sigma''$ [47]. Hence, as seen in Fig. 8(c), the standard MB equation (2) gives a poor-fitting to the normalized $\sigma''$ data in the lower frequency region. In the high-frequency region, the fitting of $\sigma''$ by equation (2) is reasonably good.

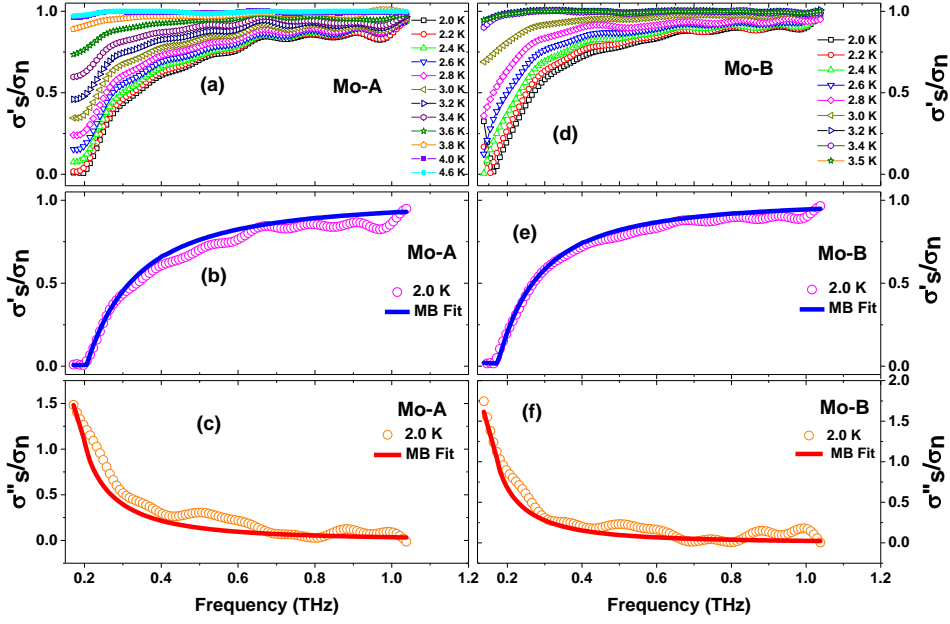

**Figure 8(a)**: Frequency dependence of the real part of the normalized THz conductivity of the Mo-A sample at different temperatures below the $1.1T_C$. A clear gap edge marked by vanishing real part of the normalized THz conductivity is seen in the film. **(b)- (c):** The normalized real and imaginary parts of the THz conductivity (open circles) of the Mo-A sample at 2 K and the fittings obtained using the MB formalism (solid lines) **(d):** Frequency dependence of the real part of the normalized THz conductivity at different temperatures for the Mo-B sample showing the signature of the gap edge. **(e)- (f):** The normalized real and imaginary parts of the THz conductivity of the Mo-B film at 2 K and the fittings obtained using the MB formalism.

The normalized plots of $\sigma'$ and $\sigma''$ for the sample, Mo-B are shown in Figs. 8(e) and (f), respectively. The curves representing the MB fit to the normalized $\sigma'$ and $\sigma''$ are also shown in the Figs. as solid lines. For the Mo-B film, the fit to the conductivity data gives the value of $2E_g$ to be 0.72 meV. It can be observed from Figs. 8(b)-(c) and (e)-(f) that the fits to the MB formalism are better for the complex conductivity of the Mo-B film as compared to Mo-A. This is probably due to slightly less disorder in the Mo-B film. None of the films shows signs of anomalous dissipation below the gap edge. This is an essential feature for the materials to be used as detectors since the dissipation below the gap edge can be detrimental to the detector performance and sensitivity [38].

The granular superconductors are made up of small metallic grains embedded in an amorphous matrix. The thickness of the inter-grain amorphous barrier determines the phase coherence among different grains. For sufficiently short coherence lengths, the sample may even show a cross over from the BCS regime to Bose-Einstein condensate comprising only one Cooper pair. In some materials, this crossover manifests itself in the increase of the strong coupling ratio [38]. Reports are available in which the coupling ratio $2\Delta/k_BT_C$ decreases monotonically even below the weak coupling limit of 3.53 with the increase in the amount of disorder [48]. It is believed that such difference arises from the difference in the nature and origin of metal to insulator transitions from being Mott type in materials showing strong coupling to that of Anderson type in systems with coupling ratio falling below the weak limit [38]. At 2 K, the coupling ratio is found to be 2.37 and 2.58 for the Mo-A and Mo-B films, respectively. These values are lower than the BCS weak coupling ratio of 3.5. Similar ratios have earlier been reported for highly disordered NbN samples having identical values of $k_Fl_e$ [48]. However, it can be noticed that unlike the case of highly disordered NbN [48] and MoN [52] films, the complex optical conductivity in our samples is reasonably reproduced by the MB formalism. In comparison, the presence of anomalous absorption below the gap edge was not seen in the highly disordered Al thin films that show strong coupling for the whole range of disorder [38].

### 3.4 Penetration depth, superfluid density, and superfluid stiffness

The complex conductivity data allows us to estimate different energy and length scales, along with the superfluid density of the superconducting state. The in-plane penetration depth ($\lambda$) as a function of temperature for the Mo-A and Mo-B films have been estimated using the $\sigma''(T)$ data for 0.2 THz using the relation: $\lambda(T) = \frac{c}{\sqrt{4\pi\omega\sigma''(T)}}$, where $c$ is the vacuum speed of light and $\omega$ is the frequency of the radiation [47]. The plots (open circles) of the $\lambda^{-2}(T)$ for the Mo-A and Mo-B films are shown in Figs. 9(a) and (b), respectively. Since the electronic mean free paths for both the samples are very small (~ 1 – 2 Å), the dirty-limit BCS expression[53,54] $\frac{\lambda^{-2}(T)}{\lambda^{-2}(0)} = \frac{\Delta(T)}{\Delta(0)}\tanh\left[\frac{\Delta(T)}{2k_BT}\right]$ fits the $\lambda^{-2}$(T) data with $\Delta(0)$ as the fitting parameter. The fit for both the thin films are shown (solid lines) in the Figs. 9 (a) and (b). The values of the $\lambda(0)$ for Mo-A and Mo-B are 721 nm and 703 nm respectively. The zero temperature superconducting gap ($\Delta(0)$) has been estimated to be 0.42 meV for the Mo-A sample and 0.38 meV for the Mo-B sample.

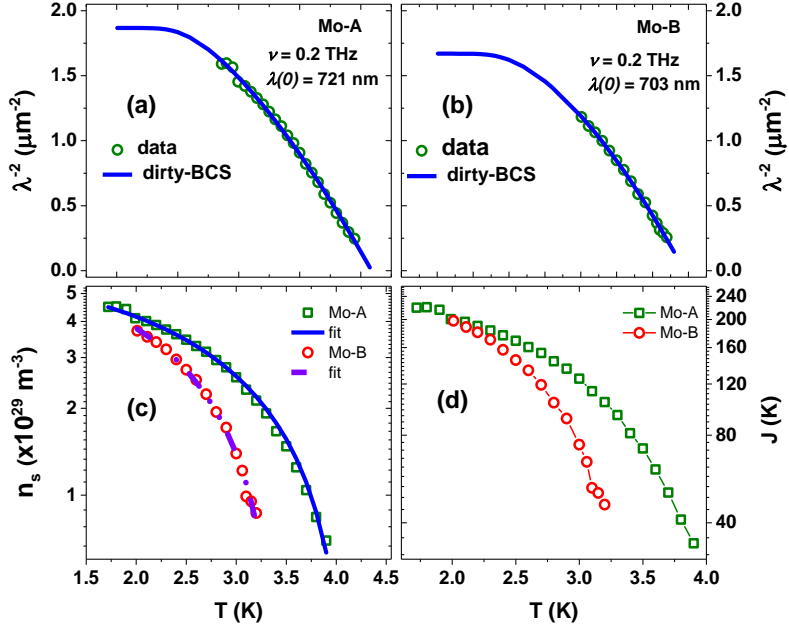

**Figure 9(a)-(b)**: Temperature dependence of $\lambda^{-2}$ for the Mo-A and Mo-B thin films. The $\lambda^{-2}(T)$ has been fitted with the dirty-limit BCS model, and the fits are shown as blue lines. **(c):** Superfluid density as a function of the temperature for the two films and the two-fluid model fits. **(d):** Temperature dependence of superfluid stiffness. The superfluid stiffness at the lowest temperature is greater than the gap for both the samples.

Using the Kramers-Kronig transformation for the superfluid contribution to the complex conductivity, the superfluid density ($n_s$) can be estimated using the relation: $n_s = \frac{2\pi m^*}{e^2} \nu \sigma''(\nu)|_{\nu=0}$, where m* is the effective mass of the quasi-particles and ν is the radiation frequency in Hz [52]. Figure 9(c) depicts the superfluid density as a function of temperature for the Mo-A and Mo-B films. The solid lines are the fit to the two-fluid model. The reliable fitting of the data with the empirical model $n_S = n_S(0)\left(1 - \left(\frac{T}{T_C}\right)^2\right)$ allows us to extract the value of $n_s(0)$ at zero temperature. The value of $n_s(0)$ for Mo-A film has been estimated to be $5.4 \times 10^{29}$ m$^{-3}$. This value agrees with the superfluid density calculated using the partial sum rule of $\sigma'$ at 2 K and the Hall effect measurements. For the film Mo-B, the value of $n_s(0)$ comes out to be $5.7 \times 10^{29}$ m$^{-3}$. Though the Mo-B film's normal carrier density is less than that of Mo-A, due to a smaller amount of disorder, the Mo-B film has a slightly larger number of Cooper pairs ($n_s/2$). This may be due to reduced scattering in the Mo-B as compared to the Mo-A film.

The temperature variation of the superfluid stiffness $J(T)$ has been estimated from the temperature dependence of penetration depth using the relation $J = 0.62\frac{d}{\lambda^2}$, where the film thickness ($d$) has been taken in Å, and $\lambda$ in $\mu$m [52,55]. The temperature variation of $n_s$ and $J$ are similar. However, the superfluid stiffness is an energy scale that gives an idea about the loss of superconductivity either via loss of pairing amplitude or phase coherence loss. Since in our case, the value of $J(2\ K)$ is 19 meV and 17 meV for the Mo-A and Mo-B films respectively, and the corresponding superconducting gaps are 0.85 meV and 0.72 meV, the superconducting transition in our films is driven by the pairing amplitude where the gap parameter $\Delta$ vanishes at the $T_C$.

The present Mo films are thus poor metals with large charge carrier density, but on the verge of localization due to the presence of disorder in their lattice. A large number of charge carriers are getting localized due to the very small mean free path, but nearly all of them are finally condensing into the Cooper pairs. The superconducting properties like enhanced $T_C$, large superfluid density, and penetration depth are correlating well with the films' morphological properties.

## 4. Conclusion

In conclusion, we have investigated the structure, morphology, DC transport, and the THz optical properties of the granular Mo thin films, which show poor metallic characteristics but are still away from the SIT. The samples have a large number of charge carriers that get localized due to the small mean free path arising out of scattering from the amorphous inter-grain regions. The samples exhibit enhancement of $T_C$ due to nanometer-sized grains. They also show nearly temperature-independent normal state transport. Though the samples are disordered, we have found a large superfluid density and a superfluid stiffness greater than the gap parameter. These indicate that the individual superconducting grains of Mo are strongly phase-coherent with each other, and the transition to the normal state is due to the loss of pairing amplitude at $T_C$. The samples' complex conductivity shows a clean gap edge that is well described within the MB formalism. Analysis of the penetration depth's temperature dependence indicates the presence of isotropic $s$-wave superconductivity in our films.

The films have shown excellent normal and superconducting properties that are at par with the materials and compounds being presently used to make microwave, terahertz, and far IR detectors. The growth of elemental thin films is simpler than the reactive sputtering of the compounds and magnetic ion implantations. We have previously shown[14] that the $T_C$ of Mo films is tunable by varying the deposition parameters during sputtering. This is of particular importance as the $T_C$ of Mo films can be tuned to meet detectors' requirements for different wavelengths. The normal state resistance is of the order of 100 $\mu\Omega$. It can be further increased with the disorder as we still have a lot of headroom ($k_F l_e$ ~ 3.8 and 4.6) before the samples reach SIT (below $k_F l_e$ ~ 1), after which quantum fluctuation may deteriorate the detector performance. The high value of normal state resistivity and sheet resistance helps in increasing the efficiency of absorption of the incident photons and a considerable variation of the kinetic inductance of the detector element. The presence of large quasi-particle density aids in achieving a large kinetic inductance that would help make more compact detectors without compromising on their performance. Thus, the granular Mo thin films are promising candidates for the development of MKIDs, comparing well with the existing competing materials.

**Author's Contributions**

All authors contributed equally to the work.

**Acknowledgments**

The authors acknowledge P. Magudapathy, Materials Science Group, IGCAR for GIXRD measurements, V. R. Reddy, UGC–DAE Consortium for Scientific Research, Indore for the X-ray reflectivity measurements, and L. S. Sharath Chandra, Free Electron Laser & Utilization Section, RRCAT, Indore for helpful discussions.

**AIP Publishing Data Sharing Policy**

The data that support the findings of this study are available from the corresponding author upon reasonable request.